\documentclass[twocolumn,showpacs,preprintnumbers,amsmath,amssymb]{revtex4}

\usepackage{graphicx}
\usepackage{dcolumn}
\usepackage{bm}

\begin{document}

\preprint{APS/123-QED}

\title{Undulation Amplitude of a Fluid Membrane \\
Surrounded by Near-Critical Binary Fluid Mixtures}

\author{Youhei Fujitani}
 \email{youhei@appi.keio.ac.jp}
\affiliation{School of Fundamental Science and Technology,
Keio University, 
Yokohama 223-8522, Japan}

\date{\today}

\begin{abstract}
We consider the thermal undulation, or shape fluctuation, of
an almost planar fluid membrane surrounded by 
the same near-critical binary fluid mixtures on both sides.
A weak preferential attraction is assumed 
between the membrane and one component of the mixture. 
We use the Gaussian free-energy functional to study the equilibrium average of
the undulation amplitude within the linear approximation with respect to the amplitude.
According to our result given by a simple analytic formula, 
the ambient near-criticality tends to suppress the undulation
of a membrane, and this suppression effect 
can overwhelm that of the bending rigidity for small wave numbers.  
Thus, the ambient near-criticality is suggested to prevent a large membrane
from becoming floppy even if the lateral tension vanishes at the equilibrium.
\end{abstract}

\maketitle

\section{\label{sec:intro}Introduction}
Amphiphilic molecules can accumulate
to form a monolayer at the interface
between two phases, working as surfactants, and can also form a bilayer 
in a one-phase solvent.   In either case, a resultant fluid membrane has
the restoring force against bending 
\cite{canh}, and its shape fluctuates at the equilibrium.  
Fluid membranes are often stacked regularly
 to form a lamellar phase because 
of the balance of their interactions, one of which is 
due to steric hindrance of undulating
membranes \cite{helf3,helf2,sorn}.  The lamellar 
structure can work as a photonic device \cite{tanak}. \\

The thermal undulation, or
shape fluctuation, of the lipid-bilayer membrane \cite{sing}
can explain the flicker phenomenon of red blood cells \cite{broc}.
When the cell is not swollen, the surface tension,
or the lateral tension, of the membrane vanishes at the equilibrium
because the membrane area is determined so that
the free energy is minimized \cite{pfeu, tanf, fritz}. 
Then, the undulation amplitude
is determined by the bending energy and becomes 
scale invariant.  This causes decrease 
in the effective bending rigidity as the membrane area is larger;
a sufficiently large membrane loses its orientation to become floppy 
\cite{helf4,helf2,sorn}.   The oil-water interface can have the same property 
when saturated by surfactants \cite{taup,cap}.  \\

It is well known that a fluid mixture shows
marked concentration fluctuation with longer
correlation length
as it approaches the demixing critical point. 
If a colloidal particle is immersed in a binary fluid mixture,
its surface usually interacts unequally
with the components.  
In a near-critical binary mixture, one component
is preferentially attracted by the surface to form 
the adsorption layer whose thickness is comparable to
the bulk correlation length \cite{indekeu, liu, hanke, JCP}.
 Dynamics of a colloidal particle
immersed in such a mixture
has been recently 
studied in theoretical aspects \cite{ofk,furu,wetdrop,visc}.  
The concentration gradient due to the adsorption layer
generates additional stress including the osmotic pressure, and
affects the flow around the particle. Accordingly, for example, the
drag coefficient deviates from the Stokes law even if the
viscosity is homogeneous in the mixture.  Being a two-dimensional droplet,
a raftlike region embedded in a binary fluid membrane can also exhibit this kind of
deviation \cite{wetraft}.
In fact, the biomembrane has several components;
the critical concentration fluctuation of the membrane  
is measured experimentally,  
with its possible biological implication suggested \cite{veach2}, and  
is studied theoretically \cite{seki}.  \\

In this paper, we consider 
a fluid membrane which is not near critical
but is surrounded by near-critical binary fluid mixtures.
The ambient near-criticality should influence 
the average of its undulation amplitude when 
one component of the mixture is preferentially attracted by the membrane.
We simplify the problem as follows to study the influence. 
The temperature is assumed to be homogeneous.
The membrane, made up of a single component, is regarded 
as a thin film fluctuating around a plane; we neglect the structure of the membrane itself
by assuming the typical radius of curvature of the undulation to be
much larger than the membrane thickness.
The semi-infinite regions on both sides of the membrane are
assumed to be occupied by incompressible
binary fluid mixtures sharing the same properties.
Far from the membrane, they are static and 
in the homogeneous phase near the demixing critical point. 
Assuming them not to be very close to the critical point,
we use the Gaussian free-energy functional.  
Our calculation is performed within the linear approximation;
sufficiently small undulation amplitude and sufficiently weak preferential attraction
are assumed.\\

Our formulation is stated in the next section; some parts are the same
as used in Ref.~\onlinecite{ofk}.  The amplitude average considered here is an
equal-time correlation at
the equilibrium and does not involve the dissipation. 
Perturbative calculations in Sec.~III yield 
a set of simultaneous equations, Eqs.~(\ref{eqn:eqforvz})--(\ref{eqn:mu2}),
which we solve in Sec.~IV by assuming the Gaussian model and the weak preferential attraction.
The results are shown in Sec.~V and is discussed in Sec.~VI.
Our study is summarized in the last section together with some outlook.

\section{\label{sec:form}Formulation}
Suppose that the binary fluid mixture consisting of
two components $A$ and $B$. We write 
$\rho_{\rm A}$ and $\rho_{\rm B}$ for
their mass densities, and
$\mu_{\rm A}$ and $\mu_{\rm B}$
for the conjugate chemical potentials.  In general, 
introducing the sum $\rho \equiv \rho_{\rm A}+\rho_{\rm B}$ and  
the difference $\varphi \equiv \rho_{\rm A}-\rho_{\rm B}$, we have
\begin{equation}
\mu_{\rm A} \delta \rho_{\rm A}+ \mu_{\rm B} \delta \rho_{\rm B}
={(\mu_{\rm A}+\mu_{\rm B})\delta\rho \over 2}
+ {(\mu_{\rm A}-\mu_{\rm B})\delta\varphi \over 2}\ ,
\label{eqn:chemwork}\end{equation}
where $\delta$ implies the infinitesimal change.
Considering that the left-hand side above 
gives a part of the infinitesimal change in the free energy,
the intensive variable conjugate to $\varphi$ is $(\mu_{\rm A}-\mu_{\rm B})/2$,
which is denoted by $\mu$.\\

The concentration difference $\varphi$ depends on 
the position $\bm{r}$ in the binary mixture.  
The $\varphi$-dependent part of the free-energy density of the mixture bulk
is assumed to be the sum of
the term independent of its gradient, denoted by $f$, 
and the term proportional to its square gradient.
This kind of free-energy density is usual in
the effective coarse-grained formulation \cite{Kadar, Onukibook}.
The contribution from the interfaces between the
membrane and the mixtures on both sides of the membrane
is simply assumed to be given by the surface integral of 
the potential $f_{{\rm s}}$ determined by the value of $\varphi$
immediately near the membrane \cite{Cahn}.
This potential represents the preferential attraction.
These assumptions enable us to write the $\varphi$-dependent part of
the free-energy functional of the mixtures as
\begin{eqnarray}&&
\int_{C^{\rm e}} d\bm{r}\ 
\left\{
f(\varphi(\bm{r}))+{1\over 2}M
\left\vert\nabla\varphi(\bm{r})\right\vert^2\right\}
\nonumber\\ &&\qquad +\int_{\partial C }dS\ 
f_{{\rm s}}(\varphi(\bm{r}))
\ .\label{eqn:glw}\end{eqnarray}
The first integral is the volume integral over the 
semi-infinite regions ($C^{\rm e}$) on both sides of the membrane,
while the second integral is the surface integral
over the interfaces ($\partial C$) on both sides.  
The coefficient $M$ is a positive constant shared by
the mixtures on both sides.
Later we will assume $f$ to be a quadratic function and $f_{\rm s}$
to be a linear function.  The free-energy functional of the mixture in general 
has a $\rho$-dependent part  
other than the part given by Eq.~(\ref{eqn:glw}),
while that of the membrane involves 
the bending rigidity and the isothermal compressibility.  \\

\begin{figure}
\includegraphics[width=8cm]{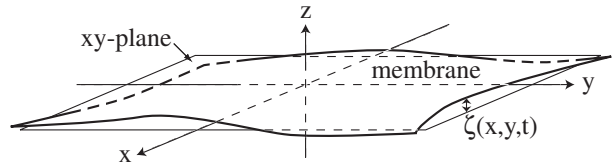}
\caption{\label{fig:memb} The fluid membrane having a single component fluctuates around the $xy$ plane.
The semi-infinite regions
on the positive and negative $z$ sides of the membrane are occupied by
binary fluid mixtures sharing the same properties. }
\end{figure}

The undulation deforms the profile of $\varphi$, and changes  
the value of Eq.~(\ref{eqn:glw}), which plays a role
of a part of the potential energy for the membrane oscillation.
This resembles the situation that
the membrane is surrounded by elastic medium \cite{rony}. 
Here, to calculate the force due to Eq.~(\ref{eqn:glw}), we need to
know how the undulation deforms the profile reversibly.
To do so,
we consider the reversible, or nondissipative, dynamics of the fluids. 
The time dependencies of $\varphi$ and
local intensive variables are thus considered below. 
The Cartesian coordinate system $(x,y,z)$ is set so that
the membrane fluctuates around the $xy$-plane (Fig.~\ref{fig:memb}).
The unit vectors along the coordinate axes are denoted by
$\bm{e}_x, \bm{e}_y$, and $\bm{e}_z$, respectively.
The $z$ coordinate of
the membrane is referred to as $\zeta$, which is a function of $(x,y)$ and the time $t$. \\

The pressure tensor of a fluid is separated in general 
into the reversible part and
the irreversible, or dissipative, part, which involves the viscosity.
For the mixture, we can obtain the former part $\Pi$
and half the local chemical potential difference $\mu$ by studying
how Eq.~({\ref{eqn:glw}) is changed by 
 an infinitesimal virtual deformation of the fluids.
This need not follow the perfect fluid dynamics.
In the bulk, the results are the same as those
in the model H, which is a standard model for the 
dynamics of a near-critical fluid \cite{Hohenberg, Onukibook},
because the same free-energy density is used \cite{ofk}. 
Below, the prime indicates the derivative with respect to the
variable. Introducing
\begin{equation}
p_{\rm osm}\equiv \varphi f'(\varphi) -f(\varphi)
\ ,\label{eqn:posmdef}
\end{equation}
which is called the osmotic pressure, and
\begin{equation}
\Pi_{\rm grad}\equiv 
-M\left({1\over 2}\vert \nabla\varphi \vert^2
+\varphi\Delta\varphi \right) \bm{ 1}+M\nabla\varphi\nabla\varphi
\ ,\label{eqn:Pigraddef}\end{equation}
where $\bm{ 1}$ denotes the isotropic tensor, we find 
\begin{equation}
\Pi={p}\bm{ 1}+p_{\rm osm}\bm{ 1}+\Pi_{\rm grad}
\ .\label{eqn:Piexpress}\end{equation} 
The scalar ${p}$ originally comes from the dependence of
the free-energy density on $\rho$.  We write $\bm{V}$
for the velocity field in the mixture.  
Assuming
$\rho$ to be a constant as in the previous
works \cite{broc,seif,physica,myjcp}, we have 
\begin{equation}
\nabla\cdot \bm{ V}=0\ .\label{eqn:incomp}
\end{equation}
We thus neglect the change in the $\rho$-dependent part 
of the free-energy density and regard ${p}$
as dependent on $\bm{ r}$ and $t$ irrespective of the local state.
  We also find 
\begin{equation}
\mu(\bm{ r},t)=f'(\varphi(\bm{ r},t))-
M\Delta\varphi(\bm{ r},t) \ ,
\label{eqn:hatmudef}
\end{equation}
which satisfies
\begin{equation}
\varphi \nabla \mu =
\nabla p_{\rm osm}+\nabla\cdot\Pi_{\rm grad}
\ .\label{eqn:inroku}
\end{equation}
We need not assume viscosity to calculate the equal-time correlation.
The dynamics of the mixture follows
\begin{equation}
\rho{\partial \bm{ V}\over \partial t}
=-\nabla {p}-\varphi \nabla\mu \ ,
\label{eqn:3Ddyn}\end{equation}
where the convective term is neglected in anticipation of the
later linear approximation.  The incompressibility condition Eq.~(\ref{eqn:incomp})
should affect the trajectory of $\varphi$. \\

Far from the membrane, 
the mixtures are assumed to be static and in the homogeneous phase, i.e., $\bm{ V}$ vanishes
and $\varphi$ is constant.  There, each of 
$\mu$, ${p}$, and $p_{\rm osm}$ is constant,
considering Eqs.~(\ref{eqn:posmdef}), (\ref{eqn:hatmudef}), and (\ref{eqn:3Ddyn}).
We assume the symmetric surroundings; the constant values
of $\varphi$, $\mu$, and ${p}$
are respectively denoted by $\varphi_\infty$, $\mu^{(0)}$, and ${p}^{(0)}$,
which are shared by the mixtures on both sides of the membrane.
The stress exerted on the membrane by
the surrounding mixture on the positive-$z$ (negative-$z$) side 
depends on $(x,y,t)$ and is denoted by $\bm{ F}^{(+)}$ ($\bm{ F}^{(-)}$).
We write $\bm{ n}$ for the unit vector which is
normal to the membrane and is directed towards the
positive-$z$ side, and define the mean curvature of the membrane $H$ 
so that its sign is positive when the center of curvature lies on the side
towards which $\bm{ n}$ is directed.\\

In Eq.~(\ref{eqn:glw}), $f_{\rm s}$ simply represents that part of the free-energy density in the bulk
which occurs only near the membrane. Thus, as $f$ generates the osmotic pressure Eq.~(\ref{eqn:posmdef}), 
$f_{\rm s}$ generates two-dimensional pressure working at
the interfaces. See Appendix A of Ref.~\onlinecite{visc} for the detail.  We have
\begin{equation}
\bm{ F}^{(\pm)}=\lim_{z\to \zeta \pm}\left\{
\mp \Pi\cdot\bm {n}
+\nabla_\parallel f_{\rm s}+2Hf_{\rm s} \bm{ n}\right\}
\ ,\label{eqn:Fplus}\end{equation}
where $\nabla_\parallel$ implies the projection of $\nabla$
on the tangent plane and
$z\to \zeta +\ (-)$ means that $z$ approaches $\zeta(x,y,t)$ with
$z-\zeta>0\ (<0)$ maintained. 
The last two terms above come from the stress due to 
the two-dimensional pressure $-f_{\rm s}$.
The boundary condition
\begin{equation}
\pm M\bm{ n}\cdot \nabla 
\varphi =f_{\rm s}'(\varphi) \quad {\rm as}\ z\to \zeta \pm
\label{eqn:phisurface}\end{equation}
should hold in the local equilibrium as well as in the global equilibrium \cite{ofk}.
The tangential components of $\bm{ F}^{(\pm)}$ vanishes;
the contribution from $M\nabla\varphi\nabla\varphi$ of Eq.~(\ref{eqn:Pigraddef})
cancels with the tangential stress due to $f_{\rm s}$, as described in Appendix D
of Ref.~\onlinecite{visc}.
In the range of $\varphi$ considered, approximating $f_{\rm s}(\varphi)$ to be
a linear function, 
we put the right-hand side (rhs) of Eq.~(\ref{eqn:phisurface}) equal to $-h$,  
where $h$ is a constant.
We can rewrite Eq.~(\ref{eqn:Fplus}) as
\begin{eqnarray}
&&\bm{ F}^{(\pm)}=\mp{h^2 \over M}\bm{ n} +\lim_{z\to \zeta \pm}
\left\{\mp {p}\pm f(\varphi)\pm {M\over 2}\vert \nabla \varphi\vert^2
\right.\nonumber\\
&&\qquad\qquad \left. \mp \mu\varphi+2Hf_{\rm s}\right\}\bm{ n}\ . 
\label{eqn:Fplus2}\end{eqnarray}
\medskip

In the previous works \cite{ofk,furu,wetraft,wetdrop,visc}, 
the diffusive flux between the two components
is considered.  Because it is proportional to
the gradient of $\mu$,
the mass conservation of each component leads to   
\begin{equation}
{\partial\varphi\over\partial t}
=-\bm{ V}\cdot \nabla\varphi+ L\Delta \mu\ ,
\label{eqn:phidyn}\end{equation}
where the Onsager coefficient $L$ is assumed to be a 
positive constant. 
Assuming that the diffusion flux cannot pass across the membrane
leads to
\begin{equation}
\bm{ n}\cdot L\nabla\mu=0\quad {\rm as}\ z\to \zeta \pm
\ .\label{eqn:musurface}\end{equation}
The diffusion should not be involved in the equal-time correlation considered here;
we will take the limit of $L\to 0+$ later.
Still, we use these two equations at this stage because, as shown later, this limit  
gives rise to the boundary layer problem, which is unfamiliar
in comparison with the problem occurring in the 
limit of zero viscosity. \\

Our calculation is performed within the linear approximation
with respect to the undulation amplitude. 
Introducing a dimensionless parameter
$\epsilon$, we define nonzero $\zeta^{(1)}$ so that we have
\begin{equation}
\zeta(\bm{x},t) =\epsilon \zeta ^{(1)}(\bm{x},t)\ .
\end{equation}
Hereafter, $\bm{x}$ represents a position on the membrane
and has coordinates $(x,y)$, in contrast with $\bm{r}$ representing
a position in the mixture. 
Up to the order of $\epsilon$,
the components of the metric tensor of the membrane
with respect to $x$ and $y$ are
the same as those of the $xy$-plane, the unit normal vector is 
\begin{equation}
\bm{n} = \bm{e}_z-{\partial \zeta\over\partial x}\bm{e}_x-{\partial \zeta\over\partial y}\bm{e}_y
\ ,\label{normal}\end{equation}
and the mean curvature is given by
\begin{equation}
H={1\over 2}\left({\partial^2 \zeta\over \partial x^2}
+{\partial^2 \zeta\over \partial y^2}\right)\ .\label{eqn:mean}
\end{equation}
We write $\bm{ v}(\bm{x}, t)$ for the velocity field of the membrane.
Assuming it to be compressible, we write
$\rho_{\rm m}(\bm{x} ,t)$ for the membrane mass per unit area, and
$p_{\rm m}(\bm{x},t)$ for its in-plane pressure field.
This field not only comes from the interaction between lipids \cite{helf2, sedo}
but can contain the interfacial tension between the membrane and the
surrounding fluid. The interfacial tension should be distinguished from 
the stress due to $f_{\rm s}$.   The former involves the density profile of the
lipids across the interface, while the latter does not.
We assume that the components of the mixture do not work as surfactants,
and thus $p_{\rm m}$ does not depend explicitly on the value of $\varphi$
immediately near the membrane.\\

The equations of motion for a viscous compressible
membrane can be found in the previous works \cite{seif,physica,myjcp,powe}.
Neglecting the membrane viscosity and using
the approximate geometrical quantities above,
we can write the momentum conservation 
in the tangential directions as
\begin{equation}
\rho_{\rm m}{\partial v_x\over\partial t}
=F_x-{\partial p_{\rm m}\over \partial x}\ {\rm and}\ 
\rho_{\rm m}{\partial v_y\over\partial t}
=F_y-{\partial p_{\rm m}\over \partial y}
\label{eqn:memx}\end{equation}
up to the order of $\epsilon$.
Here, $\bm{ F}\equiv \bm{ F}^{(+)}+\bm{ F}^{(-)}$
denotes the total stress exerted by the mixtures.
Assuming the spontaneous curvature to vanish, we 
write $c_{\rm b}H^2$ for the bending energy per unit area of the membrane, where 
 $c_{\rm b}$ is the bending rigidity \cite{canh}.
The restoring force is normal to the membrane,
and its component along $\bm{ n}$ is given by \cite{ouyang}
\begin{equation}
F_r=-c_{\rm b}\left({\partial^2\over\partial x^2}
+{\partial^2\over\partial y^2}\right)H
\ .\label{eqn:restore}\end{equation} 
Up to the order of $\epsilon$, the momentum conservation in the normal direction is
represented by
\begin{equation}
\rho_{\rm m}{\partial v_z \over \partial t}
=F_z+F_r-2Hp_{\rm m}\label{eqn:memz}
\ ,\end{equation}
while the mass conservation is represented by
\begin{eqnarray}
{\partial\rho_{\rm m}\over\partial t}=
-{\partial \rho_{\rm m} v_x
\over \partial x}-{\partial \rho_{\rm m} v_y \over \partial y}
\ .\label{eqn:memc}\end{eqnarray}
\medskip

The limit of zero viscosity in the mixture
causes the well-known boundary layer problem of the velocity field,
which we deal with by imposing the slip boundary condition
between the membrane and the inviscid fluid.   
We proceed with the calculation 
after taking this limit, and evaluate $\bm{F}^{(\pm)}$
immediately outside these boundary layers on both sides of the membrane.  
The tangential components
of the velocity need not be
continuous across the membrane,
while the normal component is continuous. 
In the limit of $L\to 0+$, as is shown later,
$\varphi$ and $\mu$ have boundary layers. 
However, at this stage,
we do not take this limit and
their spacial profiles have no rapid changes near the membrane.

\section{\label{pert}Perturbation}
In the unperturbed state ($\epsilon=0$), where the membrane is fixed on the $xy$-plane, 
$\mu$ is homogeneous over a mixture region and so is ${p}$ 
because of Eq.~(\ref{eqn:3Ddyn}) \cite{ofk}. They are respectively given by the constants
$\mu^{(0)}$ and ${p}^{(0)}$.  Up to the order of $\epsilon$, 
we expand the fields as
\begin{eqnarray}
& &\varphi(\bm{ r},t)=\varphi^{(0)}(z)+\epsilon 
\varphi^{(1)}(\bm{ r},t)\ , \nonumber\\
&& \ \mu(\bm{ r},t)=\mu^{(0)}+\epsilon \mu^{(1)}(\bm{ r},t)\ ,
\nonumber\\
& &{p}(\bm{ r},t)={p}^{(0)}+\epsilon {p}^{(1)}(\bm{ r},t)\ ,\nonumber\\
&&\ {\rm and }\ \bm{ V}(\bm{ r},t)=\epsilon \bm{ V}^{(1)}(\bm{ r},t)
\ .\label{eqn:perexp}\end{eqnarray}
On the rhs of each of these equations, 
the field with the superscript $^{(0)}$ is defined 
so that it is independent of $\epsilon$, while the field with the
superscript $^{(1)}$ is defined so that it becomes
proportional to $\epsilon$ after being multiplied by
$\epsilon$.   As shown later, 
$\varphi^{(0)}$ depends only on $z$.
For the membranous fields, 
we use similar expansions,
\begin{eqnarray}
&&\rho_{\rm m}(\bm{x},t)=\rho_{\rm m}^{(0)}+\epsilon 
\rho_{\rm m}^{(1)}(\bm{x} ,t)\ ,\nonumber\\ & &
p_{\rm m}(\bm{x},t)=p^{(0)}_{\rm m}+\epsilon p^{(1)}_{\rm m}(\bm{x}, t)\ ,\nonumber\\
&&  \bm{ v}(\bm{x},t)=\epsilon \bm{ v}^{(1)}(\bm{x},t)\ ,\nonumber\\
&&\ {\rm and}\ 
\bm{ F}(\bm{x},t)=\epsilon \bm{ F}^{(1)}(\bm{x},t)
\ ,\label{eqn:perexp2}\end{eqnarray}
where $\rho_{\rm m}^{(0)}$ and $p^{(0)}_{\rm m}$ are constants.\\

\subsection{\label{stat}Unperturbed state}
We here consider the equilibrium profile of 
$\varphi$ with the membrane fixed on the $xy$-plane;
this situation is essentially the same
as argued in Ref.~\onlinecite{Cahn}.  From Eq.~(\ref{eqn:hatmudef}),
we have
\begin{equation}
f'(\varphi^{(0)})-
M\Delta\varphi^{(0)}=\mu^{(0)}\quad {\rm for}\ z\ne 0\ .
\label{eqn:equalmu}
\end{equation}
The correlation length far from the membrane,
\begin{equation}
\xi_{{\rm c}} \equiv 
\sqrt{{M\over f''(\varphi_\infty)}}\ ,\label{eqn:zetac}
\end{equation}
is assumed to be much larger than the microscopic length, considering that
the free-energy functional Eq.~(\ref{eqn:glw}) is
a result of coarse-graining. 
Here, $f''(\varphi_\infty)$ is positive because of
the thermodynamic stability.
Equation (\ref{eqn:phisurface}) leads to
\begin{equation}
M{\partial\over\partial z}\varphi^{(0)} =\mp h
\quad{\rm as}\ z\to 0\pm \label{eqn:phisurface0}\end{equation}
Linearizing Eq.~(\ref{eqn:equalmu}) by
approximating $f'(\varphi)$ as 
$\mu^{(0)}+f''(\varphi_\infty)(\varphi-\varphi_{\infty})$, 
we obtain the equilibrium profile, 
\begin{equation}
\varphi^{(0)}(z)
=\varphi_{\infty} + 
{h\xi_{\rm c}\over M} e^{- \left\vert z\right\vert/\xi_{\rm c}}
\label{eqn:phizero}\end{equation} 
for $z\ne 0$. 
 The preferential attraction, represented by $h$, causes the concentration
difference to deviate from 
its value far from the membrane; the characteristic length is given by 
the bulk correlation length $\xi_{\rm c}$.
The approximation is valid  when
\begin{equation}
\vert hf'''(\varphi_\infty)\vert \xi_{\rm c} \ll  Mf''(\varphi_\infty)\ ,
\label{eqn:hsmall} \end{equation}
as discussed in Ref.~\onlinecite{ofk}.
We later use the Gaussian model, 
where Eq.~(\ref{eqn:phizero}) becomes exact.  \\


\subsection{\label{terms}Terms at the order of $\epsilon$}
From Eqs.~(\ref{eqn:phisurface}) and (\ref{eqn:phisurface0}), we have
\begin{equation}
\lim_{z\to 0\pm} {\partial \varphi^{(1)}\over\partial z}
=-\zeta^{(1)}\lim_{z\to 0}{\varphi^{(0)}}''(z)
\ .\label{eqn:bcphi}\end{equation}
Considering
Eq.~(\ref{eqn:musurface}), we have
\begin{equation}
L{\partial \mu^{(1)}\over \partial z}\to 0\quad {\rm as}\ z\to 0\pm 
\ .\label{eqn:musurface2}\end{equation}
Up to the order of $\epsilon$, we have
\begin{equation}
f(\varphi(\zeta+))= f(\varphi^{(0)}(\zeta+))+\epsilon \varphi^{(1)}(0+)f'(\varphi^{(0)}(0+))\ ,
\end{equation}
where $\varphi$ and $\varphi^{(1)}$ depend on $(\bm{r},t)$, $\zeta$ depends on $(\bm{x},t)$, and
$\varphi(\zeta+)$ means $\lim_{z\to \zeta+}\varphi(x,y,z,t)$, while the first term on the rhs above
equals
\begin{equation}
f(\varphi^{(0)}(0+))+\epsilon \zeta^{(1)}{\varphi^{(0)}}'(0+)f'(\varphi^{(0)}(0+))
\ .\end{equation}
Calculating similarly the other terms in Eq.~(\ref{eqn:Fplus2}), we use 
Eqs.~(\ref{eqn:equalmu}), (\ref{eqn:phizero}), and (\ref{eqn:bcphi}) 
to obtain $F_x^{(1)}=F_y^{(1)}=0$ and 
\begin{eqnarray}
&&F_z^{(1)}=\left[ -{p}^{(1)}
-\mu^{(1)}\varphi^{(0)}+\displaystyle{{h\varphi^{(1)}\over \xi_{\rm c}}}\right]_{-}
\nonumber\\ &&\qquad\quad
-{2h^2\zeta^{(1)}\over M\xi_{\rm c}}+4H^{(1)}f_{\rm s}(\varphi^{(0)}(0+)) 
\ ,\label{eqn:Fz}\end{eqnarray}
where $[\cdots]_-$ is defined as
$(\lim_{z\to 0+}\cdots)-(\lim_{z\to 0-}\cdots)$, and $H^{(1)}$ is defined as
Eq.~(\ref{eqn:mean}) with $\zeta$ replaced by $\zeta^{(1)}$.
As is mentioned at the end of Sec.~\ref{sec:form}, $z\to 0\pm$ above means
that the stress is evaluated immediately outside the boundary layers
occurring in the limit of zero viscosity.\\

In the directions of
$x$ and $y$,  we impose the periodic boundary condition, and
add an overhat  to the Fourier transform, e.g.,  
\begin{equation}
{\hat p}^{(1)}(\bm{ k},z,t)\equiv {1\over  l^2}\int_{-l/2}^{l/2}dx
\int_{-l/2}^{l/2}dy\   
p^{(1)}(\bm{x},z,t)e^{-i\bm{k}\cdot\bm{x}}
\ ,\label{eqn:four}\end{equation} 
where $\bm{ k}$ represents $(k_x,k_y)$ with
$lk_x/(2\pi)$ and $lk_y/(2\pi)$ being
integers and the period $l$ is assumed to be sufficiently large.  We have
\begin{equation}
\lim_{z\to 0\pm }{\hat V}_z^{(1)}={\hat v}_z^{(1)}= {\partial {\hat \zeta}^{(1)} \over\partial t}\ .
\label{eqn:vzl}\end{equation}
We add an overtilde to the further Fourier transform with respect to $t$,
e.g.,
\begin{equation}
{\tilde p}^{(1)}(\bm{ k},z,\omega)
={1\over 2\pi} \int_{-\infty}^\infty dt\  
{\hat p}^{(1)}(\bm{ k},z,t)
e^{i\omega t}\ .
\end{equation}
Using $k\equiv \sqrt{k_x^2+k_y^2}$, we define
\begin{equation}
{\tilde V}_\parallel\equiv 
\left(k_x{\tilde V}_x+k_y{\tilde V}_y\right)/k
\ ,\end{equation}
and define ${\tilde v}_\parallel$ similarly.
From Eq.~(\ref{eqn:3Ddyn}), we obtain
\begin{eqnarray}
&&-i\omega\rho{\tilde V}^{(1)}_\parallel
=-ik{\tilde p}^{(1)}-ik\varphi^{(0)}{\tilde\mu}^{(1)}\label{eqn:3Ddynpara}\\
&&{\rm and}\quad -i\omega\rho{\tilde V}^{(1)}_z
=-{\partial {\tilde p}^{(1)}  \over\partial z} -\varphi^{(0)}{\partial{\tilde\mu}^{(1)}\over\partial z} 
\ .\label{eqn:3Ddynz}
\end{eqnarray}
The other component
$k_x{\tilde V}^{(1)}_y-k_y{\tilde V}^{(1)}_x$ is time-invariant  
irrespective of the dynamics above in the inviscid mixture, 
and is assumed to vanish
in the calculation for the equal-time correlation.
Equation (\ref{eqn:incomp}) leads to
\begin{equation}
ik{\tilde V}^{(1)}_\parallel+{\partial {\tilde V}^{(1)}_z
\over \partial z}
=0\label{eqn:incomp2}\ .
\end{equation} 
\medskip

Deleting $\partial {\tilde p}^{(1)}/\partial z$ 
from Eq.~(\ref{eqn:3Ddynz}) and
the $z$ derivative of Eq.~(\ref{eqn:3Ddynpara}), we use
Eq.~(\ref{eqn:incomp2}) to derive
\begin{equation}
\left({\partial^2\over\partial z^2}-k^2\right) 
{\tilde V}_z^{(1)}=-{ik^2\over \rho\omega}{{\varphi}^{(0)}}'
{\tilde\mu}^{(1)}\ .\label{eqn:eqforvz}
\end{equation}
A boundary condition is given by the Fourier transform of Eq.~(\ref{eqn:vzl})
with respect to $t$.
The fields with the superscript $^{(1)}$ in Eq.~(\ref{eqn:perexp})
vanish far from the membrane.  
From Eq.~(\ref{eqn:hatmudef}), we have
\begin{equation}
\left\{ M\Delta -f''(\varphi^{(0)})\right\} \varphi^{(1)} 
=-\mu^{(1)}
\ .\label{eqn:mu1}\end{equation}
Substituting Eq.~(\ref{eqn:perexp}) into Eq.~(\ref{eqn:phidyn}), we
pick up an equation at the order of $\epsilon$.
The first term on the rhs of  Eq.~(\ref{eqn:phidyn}) generates
$-\epsilon V_z^{(1)} {\varphi^{(0)}}'$ in the resultant equation,
the Fourier transform of which gives
\begin{equation}-i\omega{\tilde {\varphi}}^{(1)}=
-{\tilde V}_z^{(1)}{\varphi^{(0)}}'+L\left({\partial^2\over\partial z^2} -k^2\right) {\tilde \mu}^{(1)}\ .\label{eqn:mu2}
\end{equation}
Let us introduce a dimensionless parameter, 
\begin{equation}
\lambda\equiv {h\xi_{\rm c}^{3/2} \over \sqrt{c_{\rm b}M}}\ .\label{eqn:lamdefi}
\end{equation}
In the next section, we
 solve the simultaneous equations, Eqs.~(\ref{eqn:eqforvz})--(\ref{eqn:mu2}),
 to calculate Eq.~(\ref{eqn:Fz}) by
introducing the Gaussian model and assuming
sufficiently weak preferential attraction to have $\lambda\ll 1$. \\

The conditions and equations
for $\mu^{(1)}$ and $\varphi^{(1)}$, given in the preceding paragraph and 
by Eqs.~(\ref{eqn:bcphi}) and (\ref{eqn:musurface2}),
are satisfied by $\mu^{(1)}=\varphi^{(1)}=0$ 
when $h$ vanishes, considering Eq.~(\ref{eqn:phizero}).  
Thus, using
$Z\equiv z/\xi_{\rm c}$, we can introduce 
dimensionless fields,
\begin{eqnarray}&&
Q\left(\bm{k}, Z, \omega\right)\equiv \displaystyle{{\xi_{\rm c}^2{\tilde \mu}^{(1)}\left(\bm{k},z,\omega\right)
\over h{\tilde \zeta}^{(1)}\left(\bm{k},\omega\right)}}\ ,\nonumber\\
&& G\left(\bm{k}, Z, \omega\right)\equiv \displaystyle{{M {\tilde \varphi}^{(1)}\left(\bm{k},z,\omega\right)
\over h{\tilde \zeta}^{(1)}\left(\bm{k},\omega\right)}} \ , \nonumber \\
&& {\rm and}\quad U\left(\bm{k}, Z, \omega\right)\equiv \displaystyle{{i{\tilde V}_z^{(1)}\left(\bm{k},z,\omega\right)
\over \omega {\tilde \zeta}^{(1)}\left(\bm{k},\omega\right)}} \ ,\label{eqn:QGU}\end{eqnarray}
which vanish far from the membrane.
Below, for conciseness, we refer to these fields as $Q(Z), G(Z),$ and $U(Z)$, respectively,
and write $\partial_Z$ and $\partial_Z^2$ for $\partial/(\partial Z)$ and $\partial^2/(\partial Z^2)$,
respectively.
We can rewrite Eq.~(\ref{eqn:eqforvz}) as
\begin{equation}
\left( \partial_Z^2-K^2 \right) U(Z)=\mp 
\lambda^2 AQ(Z) e^{\mp Z}\label{eqn:pZKU}
\end{equation}
for $\pm Z>0$, where  
$K\equiv k\xi_{\rm c}$ and
$A\equiv c_{\rm b}k^2/\left(\rho\omega^2\xi_{\rm c}^3\right)$.
With the viscosity considered,
Eq.~(\ref{eqn:pZKU}) would have some terms multiplied by
the viscosity coefficient which include
a higher derivative of $U$ with respect to $Z$.  As mentioned at the end of
Sec.~\ref{sec:form}, we consider the solution in the limit of zero viscosity,
i.e., we consider only the regions outside the
resultant boundary layers.  Thus, no more boundary layer of $U$ remains in Eq.~(\ref{eqn:pZKU}). \\ 

Because of Eq.~(\ref{eqn:vzl}), we have
\begin{equation}
U\left(\bm{k}, Z, \omega\right)\to 1\quad {\rm as}\ Z\to 0\pm\ .\end{equation}
Thus, applying the method of variation of parameters to Eq.~(\ref{eqn:pZKU}),
we obtain for $Z>0$
\begin{eqnarray}
&&U(Z)=
\left\{1-{\lambda^2 A\over 2K} \int_0^\infty d Z_1\ Q(Z_1)e^{-(K+1) Z_1}\right\}e^{-KZ}
\nonumber\\
&&\qquad -\lambda^2 A \int_0^\infty d Z_1\ \Gamma_K(Z,Z_1)Q(Z_1)e^{-Z_1}
\ ,\label{eqn:Usol1}\end{eqnarray}
where the kernel is defined as
\begin{equation}
\Gamma_K(Z,Z_1)=
-{1\over 2K} e^{-K  \left\vert Z-Z_1 \right\vert} 
\label{eqn:gamdef}\end{equation}
for $Z>0$ and $Z_1>0$.  For $Z<0$, we likewise find
\begin{eqnarray}
&&U(Z)=
\left\{1+{\lambda^2 A\over 2K} \int_{-\infty}^0 d Z_1\ Q(Z_1)e^{(K+1) Z_1}\right\}e^{KZ}
\nonumber\\
&&\qquad +\lambda^2 A \int_{-\infty}^0 d Z_1\ \Gamma_K(Z,Z_1)Q(Z_1)e^{Z_1}
\ ,\label{eqn:Usol2}\end{eqnarray}
where the kernel is also given by Eq.~(\ref{eqn:gamdef}) for $Z<0$ and $Z_1<0$. 
Thus, we have 
\begin{equation}
U(Z)=e^{-K|Z|}+{\cal O}(\lambda^2) \ ,\label{eqn:ekz}\end{equation}
where ${\cal O}(\lambda^2)$ represents the term whose quotient divided by
$\lambda^2$ does not diverge in the limit of $\lambda\to 0+$. 
Later we use the term independent of $\lambda$, $e^{-K|Z|}$, 
to calculate $Q$ and $G$.  The former result, in particular, is then
substituted into
Eq.~(\ref{eqn:Usol2}) to yield $U$ up to the order of $\lambda^3$.

\section{\label{sec:calc}Solution at $L\to 0+$}
The free-energy functional Eq.~(\ref{eqn:glw}) 
is considered as obtained after renormalized
up to the correlation length \cite{JCP}.
Assuming the correlation length to be so short
that the higher-order terms are negligible in $f$,
we use the Gaussian model,
\begin{equation}
f(\varphi)={a\over 2} \left(\varphi-\varphi_\infty\right)^2
+\mu^{(0)}\left(\varphi-\varphi_\infty\right)
\ ,\label{eqn:gauss}\end{equation}
where $a$ is a positive constant.
This constant, being the reciprocal susceptibility,
can be assumed to be proportional to the
temperature measured from the critical point.
Using Eq.~(\ref{eqn:gauss}) in Eq.~(\ref{eqn:glw}) 
amounts to assuming that the mixture is near, but
not very close to, the demixing critical point \cite{Onukibook}.
Equations (\ref{eqn:bcphi}) and (\ref{eqn:mu1}) 
respectively become
\begin{equation}
\lim_{Z\to 0\pm}\partial_Z G(Z)=-1
\label{eqn:G0}\end{equation} and
\begin{equation}
\left(\partial^2_Z -K_1^2\right) G(Z)= -Q(Z)
\ ,\label{eqn:Geq}\end{equation}
where $K_1\equiv \sqrt{K^2+1}$.
Thus, we find
\begin{eqnarray}
&&G(Z)=\left\{1+{1\over 2} \int_0^\infty dZ_1\ e^{-K_1Z_1}Q(Z_1) \right\} {e^{-K_1Z}\over K_1} 
\nonumber\\&&\qquad 
-\int_0^\infty dZ_1\ \Gamma_{K_1}(Z,Z_1) Q(Z_1) 
\label{eqn:Gsol}\end{eqnarray}
for $Z>0$, and 
\begin{eqnarray}
&&G(Z)=\left\{-1+{1\over 2} \int_{-\infty}^0 dZ_1\ e^{K_1Z_1}Q(Z_1) \right\} {e^{K_1Z}\over K_1} 
\nonumber\\&&\qquad 
-\int_{-\infty}^0 dZ_1\ \Gamma_{K_1}(Z,Z_1) Q(Z_1)  
\label{eqn:Gsolm}\end{eqnarray}
for $Z<0$.  Using ${\cal L} \equiv iLM/(\omega\xi_{\rm c}^4)$, we rewrite 
Eq.~(\ref{eqn:musurface2}) as
\begin{equation}
{\cal L} \partial_Z Q(Z)\to 0 \quad {\rm as }\ Z\to 0\pm\ .
\label{eqn:musurface3}\end{equation}
From Eq.~(\ref{eqn:mu2}), we obtain for $\pm Z >0$
\begin{equation}
{\cal L} \left(\partial_Z^2-K^2\right)
Q(Z)=\mp U(Z)e^{-|Z|}+G(Z)\ .
\label{eqn:mu3}\end{equation}

\medskip
Substituting  Eq.~(\ref{eqn:Geq}) into Eq.~(\ref{eqn:mu3}), we obtain
\begin{eqnarray}
&&-Q(Z)={\cal L} \left(\partial_Z^2-K_1^2\right) 
\left( \partial_Z^2-K^2\right)
Q(Z)\nonumber\\ &&\qquad\qquad
\pm 2Ke^{-(K+1)|Z|}+{\cal O}(\lambda^2)
\label{eqn:4kai}\end{eqnarray}
with the aid of Eq.~(\ref{eqn:ekz}). 
We are interested in the limit of $L\to 0+$, i.e., $-i{\cal L}\to 0+$, in Eq.~(\ref{eqn:4kai}),
which gives the singular perturbation problem \cite{bender}. 
For $0<|Z|\ll 1$, 
we introduce $u \equiv {\cal L}^{-1/4}Z$
and $q(u)\equiv Q({\cal L}^{1/4}u)$
to rewrite Eq.~(\ref{eqn:4kai}) as 
\begin{eqnarray}&&
-q(u)
=\left( {\partial^2\over\partial u^2}-\sqrt{{\cal L}}K_1^2\right)
\left({\partial^2\over\partial u^2}-\sqrt{{\cal L}}K^2\right) q(u)
\nonumber\\ &&\quad
\pm 2K\exp{\left\{-(K+1){\cal L}^{1/4}|u|\right\} }+{\cal O}(\lambda^2)\ ,
\label{eqn:4kai2}\end{eqnarray}
which has regular solutions of $q$ even in the limit of $-i {\cal L}\to 0+$.
The highest derivative is free from ${\cal L}$ in the above, unlike in Eq.~(\ref{eqn:4kai}).
Thus, on each side near $Z=0$, there is a boundary layer, whose thickness tends to zero as $L\to 0+$.
In this limit, considering Eq.~(\ref{eqn:4kai}),
$Q(Z)$ is asymptotically equal to
\begin{equation}
\mp 2Ke^{-(K+1)|Z|}+{\cal O}(\lambda^2)
\label{eqn:outer}\end{equation}
for $\pm Z>0$ outside the thin layers.  This outer solution
satisfies the boundary condition for $|Z|\to\infty$,   
mentioned just below Eq.~(\ref{eqn:QGU}), while it does not
satisfy Eq.~(\ref{eqn:musurface3}). 
An alternative way to Eq.~(\ref{eqn:outer}) is as follows.
The boundary layer of $Q(Z)$ yields that of $G(Z)$ because of Eq.~(\ref{eqn:Geq}).
The outer solution of $G(Z)$ is $\pm U(Z)e^{-|Z|}$ considering Eq.~(\ref{eqn:mu3}).
Substituting this into Eq.~(\ref{eqn:Geq}) gives the outer solution of $Q(Z)$, i.e., 
Eq.~(\ref{eqn:outer}), with the aid of Eq.~(\ref{eqn:ekz}).
Thus, once the boundary layers are recognized, 
Eqs.~(\ref{eqn:4kai}) and (\ref{eqn:4kai2}) are dispensable in deriving Eq.~(\ref{eqn:outer}). \\

We define $Q_{\rm in}(Z)$ so that $Q(Z)$ equals the sum of  $Q_{\rm in}(Z)$ and 
 Eq.~(\ref{eqn:outer}); $Q_{\rm in}(Z)$ rapidly becomes zero as $|Z|$ increases beyond 
the thickness of the boundary layer.
In the limit of $L\to 0+$, substituting the sum 
into Eq.~(\ref{eqn:Gsol}) gives
 \begin{eqnarray}
&&G(Z)=e^{-(K+1)Z}\nonumber\\
&&-{ e^{-K_1 Z}\over K_1} \left\{ K- \int_0^\infty dZ_1\   
 Q_{\rm in}(Z_1) \right\}+ {\cal O}(\lambda^2)
\label{eqn:new}\end{eqnarray}
for $Z>0$ outside the boundary layer.  There, we should have $G(Z)=U(Z)e^{-Z}$ from
Eq.~(\ref{eqn:mu3}).  Thus, we use Eqs.~(\ref{eqn:ekz}) to find
\begin{equation}
\lim_{L\to 0+} \int_0^\infty dZ_1\  Q_{\rm in}(Z_1)=K+{\cal O}(\lambda^2) \ ,
\label{eqn:inner}\end{equation}
and thus Eq.~(\ref{eqn:Gsol}) in the limit of $Z\to 0+$ gives
 \begin{equation}
\lim_{L\to 0+} G(0+) =1 +{\cal O}(\lambda^2)\ .
\label{eqn:G0plus}\end{equation}
This happens to be equal to the same limit of
the outer solution of $G(Z)$.
If $L$ were assumed to vanish from the beginning, 
Eq.~(\ref{eqn:musurface3}) would be trivial and
Eq.~(\ref{eqn:mu3}) would give 
$G(Z)=\pm U(Z)e^{-|Z|}$ for $\pm Z> 0$.
This overall solution contradicts with Eq.~(\ref{eqn:G0}).
This means that we cannot assume $L$ to vanish
from the beginning.  We use Eq.~(\ref{eqn:bcphi}) to derive Eq.~(\ref{eqn:Fz})
because of the statements in the last paragraph of Sec.~{\ref{sec:form}. 
Thus, we cannot take the limit of $L\to 0+$ before
taking the limit of $Z\to 0\pm$ in evaluating $\bm{F}^{(\pm)}$ in the
reversible dynamics. \\

Using Eqs.~(\ref{eqn:Usol1}), (\ref{eqn:outer}) and (\ref{eqn:inner}), we obtain 
\begin{eqnarray}
&&\lim_{L\to 0+}
\lim_{Z\to 0+}\partial_Z U(Z)\nonumber\\
&& =
-K-{\lambda^2 AK \over K+1}+\lambda^2 A \int_0^\infty dZ_1\ Q_{\rm in}(Z_1)
+{\cal O}(\lambda^4)\nonumber\\
&&=-K+{\lambda^2 AK^2\over K+1}+{\cal O}(\lambda^4)
\ .\label{eqn:partialU}\end{eqnarray}
For $Z<0$, using the procedure leading to Eq.~(\ref{eqn:inner}),  we obtain 
\begin{equation}
\lim_{L\to 0+} \int_{-\infty}^0 dZ_1\  Q_{\rm in}(Z_1)=-K+{\cal O}(\lambda^2) 
\ .\label{eqn:inner2}\end{equation}
Thus, considering Eqs.~(\ref{eqn:Usol1}), (\ref{eqn:Usol2}), 
(\ref{eqn:outer}) and (\ref{eqn:inner2}), we find $U(Z)$ to be even with respect to $Z$
 up to the order of $\lambda^3$. 
 With the aid of Eq.~(\ref{eqn:incomp2}), $V_\parallel(z)$ is found to be odd with respect to
$z$ up to this order, in spite of which $v_\parallel$ does not vanish because
 of the slip boundary condition. 
From Eqs.~(\ref{eqn:Gsol}), (\ref{eqn:Gsolm}), (\ref{eqn:outer}), (\ref{eqn:inner}) and (\ref{eqn:inner2}), 
$G(Z)$ turns out to be odd with respect to $Z$ up to the order of $\lambda$.
There are three terms on the rhs of Eq.~(\ref{eqn:Fz}).
The Fourier transform of its first term is thus found,
with the aid of Eqs.~(\ref{eqn:3Ddynpara}) and (\ref{eqn:incomp2}), to be given by
\begin{eqnarray}
&&2\lim_{z\to 0+} \left\{\left({-i\omega \rho\over k^2}\right) 
{\partial {\tilde V}^{(1)}_z\over \partial z}+
{h{\tilde \varphi}^{(1)} \over \xi_{\rm c}}\right\}\nonumber\\
&&={2c_{\rm b}{\tilde \zeta}^{(1)}\over \xi_{\rm c}^4}\lim_{Z\to 0+} \left\{-{\partial_Z U(Z)\over A}+\lambda^2G(Z)
\right\}
\label{eqn:Fz2}\end{eqnarray} up to the order of $\lambda^3$.  
Substituting Eqs.~(\ref{eqn:G0plus}) and (\ref{eqn:partialU}) into Eq.~(\ref{eqn:Fz2}), we find 
the Fourier transform of 
the sum of the first and second terms on the rhs of 
Eq.~(\ref{eqn:Fz}) in the limit of $L\to 0+$ to be  
\begin{equation} 
{2\rho\omega^2 {\tilde \zeta}^{(1)}\over k}-{2c_{\rm b}{\tilde \zeta}^{(1)}\over \xi_{\rm c}^4}
\lambda^2 d(K)
\label{eqn:Fz3}\end{equation}
up to the order of $\lambda^3$, where 
\begin{equation}
d(K)\equiv {K^2\over K+1}
\ .\label{eqn:dK}\end{equation}
Equation (\ref{eqn:Fz3}) originates from 
Eq.~(\ref{eqn:Piexpress}); the term $M\nabla\varphi\nabla\varphi$ in
Eq.~(\ref{eqn:Pigraddef}) does not contribute to this result, 
as mentioned below Eq.~(\ref{eqn:phisurface}).
Neither does the term $M\vert\nabla\varphi\vert^2/2$ in Eq.~(\ref{eqn:Fplus2});
the second term on the rhs of Eq.~(\ref{eqn:Fz}) can be traced to this term but
cancels with the term involving $G$ in Eq.~(\ref{eqn:Fz2}) because of Eq.~(\ref{eqn:G0plus}).
Which component is preferred by the membrane is not involved in deriving
Eq.~(\ref{eqn:Fz3}), which does not contain  
a term with odd powers of $h$.  
Using Eqs.~(\ref{eqn:Fz3}) and (\ref{eqn:dK}), we can evaluate $\bm{F}$
in the limit of $L\to 0+$.  \\

Introducing the isothermal compressibility of the membrane $\kappa$, we assume
$\kappa p^{(1)}_{\rm m}=\rho_{\rm m}^{(1)}/\rho_{\rm m}^{(0)}$. 
Noting the statement above Eq.~(\ref{eqn:Fz}), we
use Eqs.~(\ref{eqn:memx}) and (\ref{eqn:memc}) to obtain
\begin{equation}
{\partial \over\partial t}{\hat \rho}^{(1)}_{\rm m}
=-ik\rho^{(0)}_{\rm m}{\hat v}_\parallel^{(1)}\quad {\rm and}\quad 
{\partial \over \partial t}{\hat v}_\parallel^{(1)}
=-{ik \over \rho_{\rm m}^{(0)}\kappa} {\hat \rho}_{\rm m}^{(1)}
\ ,\label{eqn:parallel}\end{equation}
which describe the nondissipative oscillation in the tangential direction. 
 That in the normal direction,
independent of Eq.~(\ref{eqn:parallel}),
leads to the equilibrium average of the undulation amplitude, as shown in the next section.

\section{\label{sec:res}Results}
The stress  $-p^{(0)}_{\rm m}$ gives 
the lateral tension or surface tension referred to in Refs.~\onlinecite{broc, helf2, sorn, seif} and \onlinecite{sedo},
where the preferential attraction is not considered.
This equilibrium stress vanishes when the membrane is not forced to be stretched or compressed,
as is mentioned in the second paragraph of Sec.~\ref{sec:intro}.
Then, if the preferential attraction occurs in the ambient near-criticality,  
the stress defined as
\begin{eqnarray}
\sigma_{\rm l} \equiv -p^{(0)}_{\rm m}+2f_{\rm s}(\varphi^{(0)}(0+))
\label{eqn:lat}\end{eqnarray} 
vanishes, considering the statements above and below Eq.~(\ref{eqn:Fplus}).
The factor $2$ above comes from the two interfaces on both sides of the membrane.
 Thus, in more general, $\sigma_{\rm l}$ is regarded as the lateral tension at the equilibrium. \\

The Fourier transform of Eq.~(\ref{eqn:restore})
is given by $-c_{\rm b}k^4{\tilde \zeta}/2$ because of Eq.~(\ref{eqn:mean}). 
Equation (\ref{eqn:memz}) yields
\begin{equation}
-i\omega\rho^{(0)}_{\rm m}{\tilde v}_z=
{\tilde F}_z^{(1)}-\left({c_{\rm b}k^4\over 2}-p^{(0)}_{\rm m} k^2\right) {\tilde \zeta}^{(1)}
\ ,\label{eqn:memz2}
\end{equation}
which is combined with Eqs.~(\ref{eqn:Fz}) and (\ref{eqn:Fz3}) to yield
\begin{equation}
\rho^{({\rm eff})}_k {\partial{\hat v}_z^{(1)}\over\partial t}
= -\left\{{c_{\rm b} \over \xi_{\rm c}^4}  \left( {K^4 \over 2}
+2 \lambda^2 d(K) \right) +\sigma_{\rm l} k^2 \right\} {\hat \zeta}^{(1)}
\label{eqn:perpend}\end{equation} up to the order of $\lambda^3$. 
Here, we use $\rho^{({\rm eff})}_k\equiv \rho^{(0)}_{\rm m}+2\rho/k$, the second term of which
represents the induced mass \cite{com}.  
Equations (\ref{eqn:vzl}) 
and (\ref{eqn:perpend}) describe the nondissipative oscillation
in the normal direction. 
We thus find
\begin{eqnarray}&&
{l^2\epsilon^2\over 2} \sum_{\bm{ k}} \left\{ \rho^{({\rm eff})}_k
\left\vert {\hat v}_z^{(1)} (\bm{ k},t)
\right\vert^2 \right.\nonumber\\
&&  \left. + \left({c_{\rm b}k^4\over 2} + \sigma_{\rm l} k^2 + {2 h^2 \over M\xi_{\rm c}} d(k\xi_{\rm c})
\right) \left\vert {\hat \zeta}^{(1)}(\bm{ k},t)
\right\vert^2 \right\}
\label{eqn:harmony}\end{eqnarray}
to be time-independent.  As shown in the next paragraph, 
the above represents the total energy of the oscillation
in the normal direction.
Let $\langle\cdots\rangle$ indicate the equilibrium average
at the temperature $T$, and $k_{\rm B}$ denote
the Boltzmann constant.
Using the equipartition theorem, we find 
\begin{eqnarray}&&
\langle {\hat \zeta}(\bm{ k},t){\hat \zeta}(\bm{ k}',t)\rangle\nonumber\\&&\quad
= \delta_{\bm{ k},-\bm{ k}'}{k_{\rm B}T\over l^2}
\left\{ {c_{\rm b}k^4\over 2} + \sigma_{\rm l} k^2 + {2 h^2 \over M\xi_{\rm c}} d(k\xi_{\rm c})
\right\}^{-1}\label{eqn:main}
\end{eqnarray}
up to the order of $h^3$, where $d$ is defined by Eq.~(\ref{eqn:dK}). 
This is our main result; the sum in the brackets on the rhs of Eq.~(\ref{eqn:main}) is the same
as that of Eq.~(\ref{eqn:perpend}).
The average of the squared undulation 
amplitude can be calculated from Eq.~(\ref{eqn:main}) by means of
\begin{equation}
\langle \zeta(\bm{x},t)\zeta(\bm{x}, t)\rangle
= \sum_{\bm {k}} \langle {\hat \zeta}(\bm{ k},t){\hat \zeta}(-\bm{ k},t)\rangle 
\ .\label{eqn:sqave}\end{equation}

\medskip
Only in this paragraph, we suppose an external stress field exerted 
on the membrane.  We write
$\eta_z(\bm{x},t)$ for its $z$-component, which is
added to the rhs of Eq.~(\ref{eqn:memz}).
Its Fourier transform ${\hat \eta}_z(\bm{k},t)$ should appear 
on the rhs of Eq.~(\ref{eqn:perpend}) multiplied by
$\epsilon$.  Let us multiply this modified equation with $l^2 {\hat v}_z(-\bm{k},t)
=l^2\epsilon {\hat v}^{(1)}_z(-\bm{k},t)$ and
sum the resultant product over $\bm{k}$.  Then, with the aid of Eq.(\ref{eqn:vzl}), 
we find the time derivative of Eq.~(\ref{eqn:harmony})
to be given by 
\begin{eqnarray} 
&&l^2 \sum_{\bm{k}} {\hat \eta}_z(\bm{k},t){\hat v}_z(-\bm{k},t)
\nonumber\\
&&\qquad =\int_{-l/2}^{l/2}dx \int_{-l/2}^{l/2}dy
\ \eta_z(\bm{x},t)v_z(\bm{x},t) 
\ ,\end{eqnarray}
which is the work done to the membrane
per unit time by the external stress field.
This means that Eq.~(\ref{eqn:harmony}) 
is the total energy, or the effective Hamiltonian, of the oscillation in the normal direction. \\

\begin{figure}
\includegraphics[width=8cm]{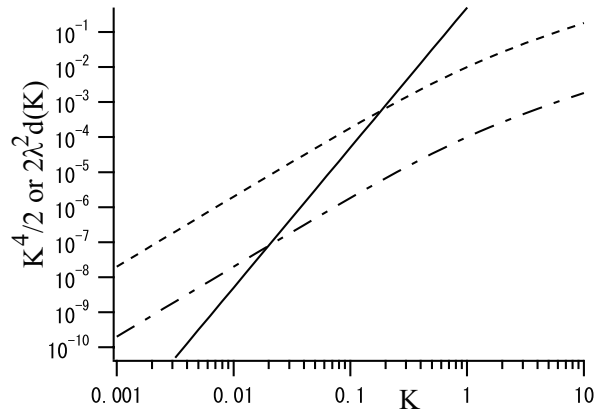}
\caption{\label{fig:amp} Logarithmic plots of the terms in the parentheses of Eq.~(\ref{eqn:perpend}).
The solid line represents the first term $K^4/2$, while the other curves represent
$2\lambda^2d(K)$. We use $\lambda=0.1$ and $0.01$ for the dashed curve and for the dash-dotted curve,
respectively. }
\end{figure}

The terms in the parentheses of Eq.~(\ref{eqn:perpend}) are plotted in
Fig.~\ref{fig:amp}.
Judging from Eq.~(\ref{eqn:dK}), $d(K)$ is positive and increases with $K=k\xi_{\rm c}$.
We have $d(K)\approx K^2$
as $K$ is small enough and $d(K)\approx K$ as $K$ is large enough, which can also be read from
the curves in Fig.~\ref{fig:amp}.
We use the hydrodynamics based on Eqs.~(\ref{eqn:glw}) and (\ref{eqn:gauss}).
Considering the statement above Eq.~(\ref{eqn:gauss}), 
our result ceases to be valid for the length scale much smaller than the correlation length,
i.e., for $K\gg 1$.  Because $d(K)$ is positive, 
the term involving $d$ in Eq.~(\ref{eqn:main}) tends to 
suppress each wave-number
component of the undulation amplitude.  
This term increases to cause more suppression 
as $k$ is larger, as $h$ is larger, and as $\xi_{\rm c}$ is larger.
The increase due to larger $\xi_{\rm c}$ is more distinct 
when $K$ is small enough to give our reliable result
$d(K)\approx K^2$ than otherwise. 
When $|h|$ and $\xi_{\rm c}$ are larger, the adsorption layer is
also remarkable in its thickness and amplitude, considering Eq.~(\ref{eqn:phizero}).  \\

\section{\label{sec:dis}Discussion}
The average of the undulation amplitude is not determined only by the equilibrium lateral tension and
bending rigidity in particular when the undulation causes a significant change in $\varphi$.
This surely occurs when near-critical
binary fluid mixtures surround the membrane with the preferential attraction 
to make the adsorption layer remarkable.  
To find the change, we use the reversible dynamics
coming from Eq.~(\ref{eqn:glw}) and the membrane energy.
We assume the fluids to be inviscid from the beginning
and then take the limit of $L\to 0+$. 
Discontinuous motion is inevitable immediately near the membrane
in these limits for the reversibility.  
The incompressibility is assumed for the surrounding fluids. 
Using all these conditions, we can determine
the reversible oscillation of 
the profile of $\varphi$ associated with that of the membrane shape $\zeta$.  \\

The equal-time correlation can be calculated generally in terms of the statics.
As shown in the Appendix, one may use Eq.~(\ref{eqn:glw}) and the homogeneous chemical potential to calculate
the probability distribution of the deviations of $\varphi$ and $\zeta$
up to their second order, and then integrate the distribution
with respect to $\varphi$ to obtain the effective Hamiltonian for $\zeta$
without using the dynamics.
However, the result is not consistent with Eq.~(\ref{eqn:main}).
The integration with respect to $\varphi$ mentioned above
 amounts to specifying the profile of $\varphi$
for a given $\zeta$, considering that the involved 
probability distribution is Gaussian.   In the oscillation of $\zeta$, the trajectory of
$\varphi$ thus determined is not necessarily the same as the
trajectory determined by the reversible dynamics, on which
the discontinuous motion and the
incompressibility impose constraints as shown in the present study.
See the Appendix for the detail.
When the membrane is surrounded by incompressible one-component fluids without $\varphi$,
we need not care about their motion in calculating the undulation amplitude 
because the motion contributes only to the kinetic energy.   
However, it is not the case in our problem, where changes of $\varphi$ and $\mu$
are correlated with the membrane motion.  Furthermore,
$\varphi$ influences the stress
exerted on the membrane and the potential-energy part, dependent on the shape $\zeta$, in the effective Hamiltonian. 
In the present study, we calculate the dependence of $\varphi$ on $\zeta$ by using the reversible dynamics,
instead of starting with the free-energy functional containing all the required
information for the statics.  \\

As mentioned in the last paragraph of the preceding section,
the ambient near-criticality tends to 
suppress the undulation amplitude more remarkably as the
the adsorption layer is more remarkable.  As far as our result remains valid,
the suppression is also more remarkable
as the wave number of the undulation is larger.
Then, like the membrane, the adsorption layer would wrinkle more severely.
If there is no preferential attraction, 
Eq.~(\ref{eqn:main}) is reduced to the previous result \cite{broc, helf2, sorn}.
Then, Eq.~(\ref{eqn:sqave})
can be calculated as
\begin{equation}
{k_{\rm B} T\over (2\pi)^2} \int_{2\pi/l}^{2\pi/s} dk\ 2\pi k \left( {c_{\rm b}k^4\over 2}-p^{(0)}_{\rm m} k^2\right)^{-1}
\ ,\label{eqn:integ}\end{equation} 
where $s$ is the lower cut-off length.
For the membrane suspended freely in a fluid,
we neglect $p^{(0)}_{\rm m}$ to obtain
\begin{equation}
\langle \zeta(\bm{x},t)\zeta(\bm{x}, t)\rangle={k_{\rm B} T\over 2\pi c_{\rm b}} \left( {l\over 2\pi}\right)^2
\ ,\end{equation}
which implies that the averaged undulation amplitude is scale invariant \cite{helf2}.
This is derived by the $k^{-3}$ dependence of the integrand 
of Eq.~(\ref{eqn:integ}), which dependence comes from the bending energy.    
If the preferential attraction occurs to give $h\ne 0$
and if the equilibrium lateral tension vanishes to give $\sigma_{\rm l}=0$,
the last term $\propto d(K)\propto k^2$ can overwhelm the term $c_{\rm b}k^4/2$ in the brackets of
Eq.~(\ref{eqn:main}) for small values of $k$, or equivalently
the second term can overwhelm the first term in the parentheses of Eq.~(\ref{eqn:perpend})
for small values of $K$.  In Fig.~\ref{fig:amp}, each of the curves is above the line for small values of $K$.
When $K$ is smaller than the value at the intersection,
the term due to the ambient near-criticality combined with the
preferential attraction overwhelms the term due to the bending rigidity.
The intersection occurs at a smaller value of $K$ as $\lambda$ is smaller, and in particular occurs
at sufficiently small $K$ for $\lambda\ll 1$.
\\

Around the room temperature, for example,
aqueous solutions of
2-methyl propanoic acid and 
1-propoxy 2-propanol respectively have the upper and
lower consolute points \cite{liqliq}.  It is probable, however, that
the structure of the lipid-bilayer membrane is disordered
when it is immersed in either of these solutions, considering that it has the 
affinity to alcohol \cite{alc}.  Thus, these solutions
would not be available for experimental check of our result.
In the coacervation of 
aqueous solutions of elastin-related
polypeptides, the lower consolute points
are around the room temperature \cite{kaibara}.
The vesicle made of the lipid-bilayer membrane 
can contain polyethylene glycol and
dextran aqueous solution \cite{dimova}, which has the demixing critical pont
around the room temperature \cite{peg}.
Our result may be observed 
in either of these polymer solutions if not blurred by the polymer dynamics.
Sodium dodecyl-sulphate (SDS), water and pentanol form a
lamellar phase with dodecane being the solvent \cite{bel}.
Adding some fluorocarbon to the solvent,
we may check our result experimentally,
considering that perfluoroheptane and isooctane have
the upper consolute point around the room temperature \cite{brad}. \\

For the membrane of SDS, pentanol, and water,  we have $c_{\rm b}/2=
2.1k_{\rm B} T\approx 10^{-20}$J according to previous experimental studies \cite{lei}.
The coefficient of the square gradient term, sometimes called the
influence parameter, is related to the direct
correlation function \cite{rowl}, and is linked with the interfacial tension
in the two-phase region \cite{poser}.
The parameter can be defined in general for each pair of the components,
$A$-$A$, $B$-$B$, and $A$-$B$.  The parameter of the last pair
can be regarded roughly as the geometric mean of the parameters of the 
first two pairs \cite{aiche}. We can write Eq.~(\ref{eqn:glw}) in terms of $\varphi$
by assuming negligibly small compressibility of the binary mixture.
 We cannot find out the data for the influence parameters
of perfluoroheptane and isooctane (or dodecane), but can obtain an
estimate for $M$ of their mixture or a similar mixture from the data for the pure fluid of alkane.
Its influence parameter is larger with the number of carbons per alkane molecule
\cite{kahl}, and is $10^{-16}$m$^7$/(s$^2$kg) for decane \cite{vdwexp}.
Using the Gaussian model in our formulation, we
can neglect the weak power dependence of $M$ on the correlation length
\cite{Onukibook}.  \\

The interval between SDS molecules
in the membrane of SDS, pentanol, and water is approximately $1$nm \cite{lei}.
The second integral of Eq.~(\ref{eqn:glw}) 
may be attributed to the hydrogen bonding if it is involved \cite{liu}.
Its energy is typically around $k_{\rm B}T$.  
The mass density of a mixture of perfluoroheptane and isooctane (or dodecane)
is roughly $1$g/cm$^3$. 
Using these values, for this mixture or a similar mixture we have an estimate $h\approx 10^{-6}$m$^3$/s$^2$,
which may be overestimated because the mixture and membrane would not
involve the hydrogen bonding.
Writing $T_{\rm c}$ for the critical temperature,
we have $\xi_{\rm c}\approx 3$nm at $T-T_{\rm c}=6$K
for the critical mixture of trimethylpentane and
perfluoroheptane \cite{brag}.  Substituting these values into Eq.~(\ref{eqn:lamdefi}), we estimate 
$\lambda$ to be $10^{-1}$ or smaller for  the membrane of SDS, pentanol, and water
in a mixture of perfluoroheptane and isooctane (or dodecane).
Let us next estimate $\lambda$ similarly by supposing the lipid-bilayer membrane in an aqueous solution.
An estimate of $M$ can be obtained from the data of the influence parameter for water \cite{vdwexp}
and is about $10^2$ times larger 
than the estimate for decane in units of m$^7$/(s$^2$kg) because of the smaller molecular weight. 
For the lipid-bilayer membrane, 
the bending rigidity is $10^{-19}$J \cite{duwe},
and the interval of lipid molecules is a little smaller than that of the SDS molecules
mentioned above \cite{Ben}.
These estimates lead to $\lambda=10^{-2}$ for $\xi_{\rm c}=3$nm. \\

As mentioned in the fourth paragraph of Sec.~\ref{sec:intro}, for our formulation to be valid, 
$k^{-1}$ should be much larger than the membrane thickness,
which is $4$-$5$nm for the membranes considered above \cite{lei,Ben}.
Judging from the values of $K$ at the intersections in Fig.~\ref{fig:amp},
the term due to the ambient near-criticality
becomes larger than the term due to the bending rigidity in Eqs.~(\ref{eqn:perpend}) and (\ref{eqn:main})
when $k^{-1}$ is larger than about $1.5\times 10$nm 
($1.5\times 10^2$nm) for $\lambda=10^{-1}$ ($10^{-2}$).  
It remains to be studied, however, whether the assumption of the weak
preferential attraction is valid for these values of $k$ and $\lambda$.

\section{\label{sec:sum}Summary and Outlook}
In this paper, we consider the undulation amplitude of
a fluid membrane immersed in a near-critical binary fluid mixture.
The preferential attraction, represented by
the second integral of Eq.~(\ref{eqn:glw}), causes the
adsorption layer, which is remarkable because of the near-criticality.
The additional force is generated by the resultant gradient of 
the order parameter, $\nabla\varphi$.  
The ambient mixture is not a simple bath having a homogeneous
chemical potential independent of the membrane motion.
Our problem is simplified as mentioned in the fourth paragraph of Sec.~\ref{sec:intro}.
Within the linear approximation with respect to the undulation amplitude,
we arrive at a set of simultaneous equations given by Eqs.~(\ref{eqn:eqforvz})--(\ref{eqn:mu2}).
We solve them by assuming the Gaussian model and $|\lambda|\ll 1$
in Sec.~\ref{sec:calc}.  See Eq.~(\ref{eqn:lamdefi}) for the definition of $\lambda$.
These assumptions break down
and numerical procedure would be required to solve the
equations if we consider longer correlation length,
stronger preferential attraction, and
smaller influence parameter.  \\

After calculations in Sec.~\ref{sec:calc}, we find that 
the restoring force is given by the sum in the brackets
of Eq.~(\ref{eqn:perpend}) \cite{com2}. This leads to
Eq.~(\ref{eqn:main}), which gives the mean squared amplitude for
each wavenumber.  See the last paragraph of Sec.~\ref{sec:res} for detailed discussion.
We thus find that
the ambient near-criticality combined with the preferential attraction 
tends to suppress the undulation.
A large membrane may be prevented from becoming floppy
even if the lateral tension vanishes at the equilibrium.
Possible experimental setup to check our results is discussed
in the fourth paragraph of the preceding section.
The profile of $\varphi$ near a surface is measured by means of
the reflectivity and the ellipsometry \cite{liu}.  The profile contains
a factor $h\xi_{\rm c}/M$ in Eq.~(\ref{eqn:phizero})  \cite{est}.  
The term due to the ambient near-criticality in Eq.~(\ref{eqn:main})
has another factor $2h^2/(M\xi_{\rm c})$.  Thus, 
we may obtain values of $M$ and $h$ by measuring the profile of $\varphi$ near 
a membrane fixed on some substrate and the
thermal undulation of  the membrane suspended in a near-critical binary fluid mixture.  \\

Our theory presupposes that the semi-infinite regions on both sides of the membrane
are occupied by fluids sharing
the same properties.  This presupposition of symmetric surroundings
should be given up in considering 
the surfactant monolayer at the oil-water interface. 
The near-criticality on one side can also be expected to suppress
the undulation amplitude, which remains to be studied.
It is interesting to calculate how
the interval between stacked membranes is changed by the near-criticality
of the intercalated fluids. To do this, we should study the finite-size effect
of the surrounding fluids by also considering
the interaction between membranes due to the critical adsorption \cite{JCP}.   
This line of study may suggest realization of a photonic device responding to
small temperature change.

\begin{acknowledgments}
The author thanks Dr.~R. Okamoto for helpful discussion.
Professor T. Kato 
kindly informed the author of Refs.~\onlinecite{bel}
and \onlinecite{brad}.  This work was partly financed by Keio Gakuji Shinko Shikin.
\end{acknowledgments}

\appendix*
\section{\label{app:a} A spurious way of calculating with the $\mu$-$\zeta$ correlation neglected}
Applying the equilibrium statistical physics to our system naively,
one may expect the following procedure to be another way to Eq.~(\ref{eqn:main}).
However, as mentioned in the second paragraph of Sec.~\ref{sec:dis},
it cannot be an alternative when
the fluctuations of the chemical potential and membrane shape
are correlated through the preferential attraction.
To clarify this claim, we below show the naive procedure explicitly.  \\

Equation (\ref{eqn:glw}) is a functional dependent on $\varphi$ and $\zeta$;
we add the subscript $\zeta$ to $C^{\rm e}$ and $\partial C$
to specify the regions for a given membrane shape $\zeta$.  
Apart from the membrane energy independent of $\varphi$,
if the equilibrium property were determined only by
Eq.~(\ref{eqn:glw}), the probability distribution of
$\varphi$ and $\zeta$ would be proportional to
the exponential function of  the quotient of 
\begin{equation}
\int_{C^{\rm e}_\zeta} d\bm{r}\ 
{\check f}(\varphi, \nabla \varphi)+\int_{\partial C_\zeta }dS\ 
f_{{\rm s}}(\varphi)
\label{eqn:glw2}
\end{equation}
divided by $-k_{\rm B} T$, 
where ${\check f}$ is defined as the difference of $\mu^{(0)}\varphi$
subtracted from
the integrand of the first integral in Eq.~(\ref{eqn:glw}).
We use Eq.~(\ref{eqn:gauss}) and regard $f_{\rm s}$ as the linear function
mentioned above Eq.~(\ref{eqn:Fplus2}).  
The deviation of $\zeta$ from zero and 
that of $\varphi$ from $\varphi^{(0)}$, denoted by $\varphi_1$,
cause the deviation of Eq.~(\ref{eqn:glw2}).
We obtain this deviation, denoted by $\delta\Omega$, by subtracting 
\begin{equation}
\int_{C^{\rm e}_0} d\bm{r}\ 
{\check f}(\varphi^{(0)}, \nabla \varphi^{(0)})+
\int_{\partial C_0 }dS\ 
f_{{\rm s}}(\varphi^{(0)})
\label{eqn:glw20}
\end{equation}
from Eq.~(\ref{eqn:glw2}).
If necessary to clarify the description, the superscript $^\uparrow$ ($^\downarrow$)
is added to a quantity and the region of the surrounding fluid on the positive-$z$ (negative-$z$) side.  \\

We rewrite the first term of Eq.~(\ref{eqn:glw20}) as the sum of
\begin{equation}
\int_{C^{{\rm e}\uparrow}_\zeta} d\bm{r}\ 
{\check f}(\varphi^{(0)\uparrow}, \nabla \varphi^{(0)\uparrow})
+\int_{C^{{\rm e}\downarrow}_\zeta} d\bm{r}\ 
{\check f}(\varphi^{(0)\downarrow}, \nabla \varphi^{(0)\downarrow})
\label{eqn:up}\end{equation}
and the integral of
\begin{equation}
\int_0^{\zeta(x,y)}dz\ \left\{{\check f}(\varphi^{(0)\uparrow}, \nabla \varphi^{(0)\uparrow})-
 {\check f}(\varphi^{(0)\downarrow}, \nabla \varphi^{(0)\downarrow}) \right\}
\label{eqn:zetaint}
\end{equation}
with respect to $x$ and $y$ over the region considered in Eq.~(\ref{eqn:four}).
Thus, subtracting Eq.~(\ref{eqn:glw20}) from Eq.~(\ref{eqn:glw2}),
we encounter a term
\begin{equation}
\int_{C^{{\rm e}\uparrow}_\zeta} d\bm{r}\ \left\{
{\check f}(\varphi^{\uparrow}, \nabla \varphi^{\uparrow})-
{\check f}(\varphi^{(0)\uparrow}, \nabla \varphi^{(0)\uparrow})\right\}
\ ,\label{eqn:encounter}\end{equation}
the integrand of which is rewritten as
\begin{equation}
{1\over 2}\left( a\varphi_1^{\uparrow 2}+M|\nabla\varphi^{\uparrow}_1|^2\right)
+M\nabla\cdot\left(\varphi_1^{\uparrow}\nabla\varphi^{(0)\uparrow}\right)
\end{equation}
with the aid of Eq.~(\ref{eqn:equalmu}). 
Up to the order of $\zeta^2$, Eq.~(\ref{eqn:zetaint}) equals
$-2h^2\zeta^2/(M\xi_{\rm c})$ 
because of Eq.~(\ref{eqn:phizero}). 
The second term of Eq.~(\ref{eqn:glw2})
is rewritten as the integral of
\begin{equation}
\sqrt{1+\left({\partial\zeta\over\partial x}\right)^2+ \left({\partial\zeta\over\partial y}\right)^2}
\left\{f_{\rm s}(\varphi^\uparrow(\zeta))+f_{\rm s}(\varphi^\downarrow(\zeta))\right\}
\end{equation}
with respect to $x$ and $y$ over the region considered in Eq.~(\ref{eqn:four}).
In the above, we have 
\begin{eqnarray}&&
f_{\rm s}(\varphi^\uparrow(\zeta))=f_{\rm s}(\varphi^{(0)\uparrow}(0+))-h \varphi^{\uparrow}_1(\zeta)
\nonumber\\&&
-h\left\{ \zeta{\varphi^{(0)\uparrow}}'(0+)+{\zeta^2\over 2} {\varphi^{(0)\uparrow}}''(0+)\right\}
\end{eqnarray}
with the higher-order terms neglected. Hence,
we use Eqs.~(\ref{eqn:equalmu}) and (\ref{eqn:phisurface0}) to obtain
\begin{eqnarray}
&&\delta\Omega[\varphi_1,\zeta]=\int_{C^{\rm e}_\zeta}d\bm{r}\ 
{1\over 2}\left( a\varphi_1^{2}+M|\nabla\varphi_1|^2\right)\nonumber\\
&&\quad +\int_{-l/2}^{l/2} dx\int_{-l/2}^{l/2} dy\  \left[{h\zeta^2\over M\xi_{\rm c} }
-{h\zeta\over\xi_{\rm c}}\left\{ \varphi_1^\uparrow(\zeta)-\varphi_1^\downarrow(\zeta)
\right\}\right.\nonumber\\&&\qquad\qquad
\left.+\left\{ \left({\partial\zeta\over\partial x}\right)^2+ \left({\partial\zeta\over\partial y}\right)^2
\right\} f_{\rm s}({\varphi}^{(0)}(0+)) \right]
\label{eqn:deltaomega}\end{eqnarray}
up to the second order with respect to $\varphi_1$ and $\zeta$.\\

\medskip
Let us minimize Eq.~(\ref{eqn:deltaomega}) with 
$\zeta$ fixed.  The stationary condition gives
\begin{eqnarray}&&\left(a-M\Delta\right) \varphi_1=0 \quad{\rm for}\ z\ne\zeta\quad {\rm and}
\nonumber\\&&M\bm{n}\cdot \nabla  \varphi_1=-{h\zeta\over\xi_{\rm c}}\quad {\rm at}\ z=\zeta
\ .\end{eqnarray}
Let $\phi$ denote $\varphi_1$ satisfying the above and $\varphi_1\to 0$ as $|z|\to \infty$.
We can obtain $\phi$ by using Eqs.~(\ref{eqn:G0})-(\ref{eqn:Gsolm}) with
$Q$ put equal to zero, i.e., with the chemical potential being homogeneous.  
The Fourier transform of $\phi^\uparrow$ 
with respect to $x$ and $y$ is given by
\begin{equation}
{\tilde\phi}^\uparrow (\bm{k},z)={h{\tilde \zeta}(\bm{k})\over MK_1}e^{-K_1z/\xi_{\rm c}}
\end{equation}
up to the order of $\zeta$.  
The corresponding result for ${\tilde \phi}^\downarrow(\bm{k},z)$
coincides with $-{\tilde\phi}^\uparrow (\bm{k},-z)$.  Thus, $\phi$ is different from $\epsilon\varphi^{(1)}$ in the text.
We integrate the probability distribution of $\zeta$ and $\varphi_1$
with respect to $\varphi_1$ to obtain that of $\zeta$. 
The result is proportional to the minimum of the former distribution 
with $\zeta$ being fixed because the distribution is Gaussian.
Thus, if the equilibrium property were determined only by
Eq.~(\ref{eqn:glw}), the probability distribution functional of 
$\zeta$ would be proportional to $e^{-\delta\Omega[\phi,\zeta]/(k_B T)}$, where
$\delta\Omega[\phi,\zeta]$ is found to be
\begin{equation}
l^2\sum_{\bm{k}}{\tilde\zeta}(\bm{k}){\tilde\zeta}(-\bm{k})
\left\{{h^2 \over M\xi_{\rm c}}{\check d}(K)  +f_{\rm s}(\varphi^{(0)}(0+))k^2\right\}\ ,
\label{eqn:appdO}\end{equation} where we use 
${\check d}(K)\equiv 1-K_1^{-1}$.
We can compare  the potential-energy part in Eq.~(\ref{eqn:harmony}) with
Eq.~(\ref{eqn:appdO}) after
the terms from the bending energy and the membrane pressure are supplemented.
Thus, $d(K)$ in the procedure of the text is replaced by ${\check d}(K)$ 
according to the procedure of this appendix, which is inappropriate for the reason stated in the
second paragraph of Sec.~\ref{sec:dis}.    We have
${\check d}(K)\approx K^2/2$ for $K\ll 1$, and thus even the approximate expressions for small $K$
are different between $d(K)$ and ${\check d}(K)$. 
Hence, under a given $\zeta$, the profiles of $\varphi$ 
in the two procedures should be totally different. \\


\begin{thebibliography}{99}
\bibitem{com}  Without the membrane, Eq.~(\ref{eqn:perpend})
gives the same dispersion relation as that of the capillary wave
at the water surface, $\omega\propto k^{3/2}$.
\bibitem{com2} The factor $h^2/M$ in Eq.~(\ref{eqn:main}) 
has the same dimension as the bulk pressure and the reciprocal of the bulk compressibility.
\bibitem{est} The estimates in Sec.~\ref{sec:dis} lead to $h\xi_{\rm c}/M\approx 10^{-2}$g/cm$^3$, which is
much smaller than the density of the mixture, as it should be.
\end{thebibliography}
\end{document}